\documentclass[12pt]{spieman}
\usepackage{amsmath,amsfonts,amssymb}
\usepackage{graphicx}
\usepackage{setspace}
\usepackage{tocloft}
\usepackage{journals}
\title{Synergy between \textit{Roman} and PRIMA imaging capabilities for large extragalactic surveys}

\author[a, *]{Médéric Boquien}
\author[b]{Laure Ciesla}
\author[b]{Roman Amestoy}
\author[c]{Jason Glenn}
\author[d]{Carlotta Gruppioni}
\author[e]{John-David Smith}
\affil[a]{Université Côte d'Azur, Observatoire de la Côte d'Azur, CNRS, Laboratoire Lagrange, 06000, Nice, France}
\affil[b]{Aix Marseille Univ, CNRS, CNES, LAM, 13388 Marseille, France}
\affil[c]{NASA Goddard Space Flight Center, Code 665, Greenbelt 20771, Maryland, United States}
\affil[d]{Istituto Nazionale di Astrofisica - Osservatorio di Astrofisica e Scienza dello Spazio, via Gobetti 93/3, Bologna,
Italy, I-40129}
\affil[e]{Ritter Astrophysical Research Center, University of Toledo, Toledo, OH, 43606, USA}

\cftpagenumbersoff{figure}
\cftpagenumbersoff{table} 
\begin{document} 
\maketitle

\begin{abstract}
The \textit{Roman} space telescope will be instrumental for characterizing the physical properties of galaxies and understanding their evolution across time. However, a complete view of galaxies star formation activity will only be possible with the addition of far-infrared observations that a telescope like PRIMA will be able to provide. Indeed, PRIMA's far-infrared camera will be highly sensitive to dust emission while \textit{Roman} will probe the stellar emission in rest-frame optical and ultraviolet of distant galaxies. Our aim here is to evaluate the advantage of combining large PRIMA and \textit{Roman} extragalactic surveys to retrieve the physical properties of galaxies, and compare with what we would obtain using either dataset separately. To do so, we use the CIGALE photometric modeling code to generate a far-ultraviolet to far-infrared synthetic set of dusty star-forming galaxies at redshifts from 1.5 to 2.5, simulating the observations from the main extragalactic surveys of PRIMA and \textit{Roman}. We find that the PRIMA+\textit{Roman} observations can reliably retrieve the star-formation rate (SFR), stellar masses, and dust luminosity.
\end{abstract}


{\noindent \footnotesize\textbf{*}Médéric Boquien, \linkable{mederic.boquien@oca.eu} }

\begin{spacing}{1}

\section{Introduction\label{sec:intro}}
Measuring the physical properties of galaxies is fundamental for understanding the processes that govern their evolution, one of the key open questions in extragalactic astrophysics\cite{madau2014a}. Because the electromagnetic emission of galaxies is the complex outcome of the various physical processes that drive the formation and evolution of galaxies\cite{conroy2013a}, such as star formation and black-hole growth, multi-wavelength surveys from the rest-frame far-ultraviolet (FUV) to the far-infrared (FIR) have been key in this endeavor\cite{giavalisco2004a, scoville2007a, franceschini2010a, oliver2012a, leroy2019a, davies2021a, casey2023a}.

In star-forming galaxies, the rest-frame FUV domain is dominated by the photospheric emission from massive stars that live up to a few hundred million years\cite{kennicutt1998a, kennicutt2012a, leitherer1999a}. This makes the FUV a reasonably good tracer of the recent Star Formation Rate (SFR) of galaxies, that is the pace at which its gas reservoir is transformed into stars. In effect, the relation between the SFR and other fundamental properties, such as the metallicity or the stellar mass, have become key diagnostics to constrain models of galaxy evolution\cite{daddi2007a, noeske2007a, laralopez2010a, mannucci2010a, lilly2013a}.

Yet, galaxies contain dust, which has the effect of both dimming and reddening galaxies\cite{salim2020a}. Even though the attenuation from dust, that is the absorption and the scattering out and into the line-of-sight, is maximal in the FUV, its effect is not to be neglected in the optical and the near-infrared (NIR)\cite{lofaro2017a}. Correcting for the presence of dust is therefore of importance to avoid biases in the measurement of the physical properties of galaxies.

Over the past two decades, the launch and exploitation of a broad range of space instruments has opened the rest-frame UV window across the entire universe. Likewise, highly successful IR missions from IRAS\cite{neugebauer1984a} to the \textit{James Webb} Space Telescope (JWST) have provided us with a superb view of dust emission\cite{kennicutt2003a, kennicutt2011a, lee2023a} from space, whereas ground-based (sub-)millimeter radio observatories such as the Atacama Large Millimeter/submillimeter Array, the James Clerk Maxwell Telescope, and the Large Millimeter Telescope are sensitive to dust emission of more distant galaxies\cite{simpson2015a, zavala2015a, lefevre2020a, bouwens2022a} as the atmosphere is opaque to observed-frame FIR radiation. However, these surveys have suffered from limited sensitivity, spatial resolution, or wavelength coverage, restricting the studies of dust to somewhat lower redshifts and/or brighter objects.

In substance, the current lack of FIR instrumentation in space means that, at the moment, we have no window on the FIR emission of nearby galaxies and also on the MIR for more distant objects. One of the most promising initiatives for addressing some of these limitations to observe the dust in distant galaxies is the PRIMAger\cite{Ciesla25} IR camera on-board the PRobe far-Infrared Mission for Astrophysics\cite{Glenn25} (PRIMA). In this article, we develop a simple straightforward proof-of-concept showing that PRIMA and the \textit{Nancy Grace Roman} Space Telescope (\textit{Roman}), which probes the optical and NIR domains, are synergistic. In combination, they offer both the sensitivity and multiwavelength coverage necessary to enable exquisite measurements of the physical properties of galaxies over unprecedented large samples of galaxies thanks to their efficient mapping capabilities. We focus in particular on four key physical properties, two related to stellar populations (the stellar mass and the star formation rate), and two related to the dust (the total dust luminosity and the relative abundance of Polycyclic Aromatic Hydrocarbon).

This article is structured as follows. We first introduce the PRIMA and \textit{Roman} missions in Sec.~\ref{sec:next-gen} and lay out the methodology in Sec.~\ref{sec:methodology}. We present the results and discuss them in Sec.~\ref{sec:results}. Finally, we conclude in Sec.~\ref{sec:conclusion}. In this article, we assume a Planck 2018\cite{planck2020a} $\Lambda$-Cold Dark Matter cosmology ($H_0=67.66$~km~s$^{-1}$~Mpc$^{-1}$, $\Omega_m=0.30966$) as well as a Chabrier\cite{chabrier2003a} initial mass function (IMF).

\section{Next-generation space-borne instruments\label{sec:next-gen}}
\subsection{PRIMA\label{sec:intro_prima}}
With the demise of \textit{Herschel}, our window on the far-IR emission has been severely restricted for galaxies beyond the nearby universe. Even though JWST is now providing us with an exquisite view of galaxies, it is limited to the mid-IR in the observed frame, limiting its ability to probe dust in distant galaxies, and its small field of view makes it poorly adapted for large surveys. The most promising project to reopen the window on dust emission in galaxies beyond the nearby universe is the PRIMA FIR observatory. One of the key objectives of PRIMA will be to measure the evolution of galaxies and their ISM across the universe\cite{Glenn25}.

To achieve its goals, PRIMA is designed to be a 1.8~m telescope cryogenically cooled to 4.5~K with two instruments on board:
\begin{itemize}
    \item FIRESS\cite{Bradford25}, a spectrometer covering the 24-235\,$\mu$m wavelength range in 4 grating modules with $R>85$. The high-res mode gives R of thousands across full band.
    \item PRIMAger\cite{Ciesla25}, a FIR camera offering two imaging modes: the Hyperspectral mode will cover the 24-84\,$\mu$m wavelength range with a spectral resolution $R\sim10$, while the Polarimetric mode will have four broad-band filters, sensitive to linear polarization, from 80 to 261\,$\mu$m.
\end{itemize}

Different surveys designs are envisioned for PRIMAger such as a multi-tiered photometric survey with a total observing time of 1500\,h, distributed over 1\,deg$^2$ for the Deep survey and 10\,deg$^2$ for the Wide survey.
They will detect and study galaxies hundreds of times fainter than were reachable by previous FIR observatories (e.g. \textit{Herschel}, \textit{Spitzer}), reaching the
bulk of the population to cosmic noon and beyond.

\subsection{\textit{Roman} Space Telescope}\label{sec:intro_roman}
While PRIMA/PRIMAger has great potential for discovery on its own, as mentioned in Sec.~\ref{sec:intro}, the combination with shorter wavelength observations is expected to greatly enhance its capabilities for measuring the physical properties of galaxies, from their SFR to their stellar mass. Various new and planned projects would offer systematic surveys of a large fraction of the sky from European Space Agency's Euclid mission to NASA's \textit{Roman} observatory. Identified as the top-priority space mission for the 2020s in the 2010 Decadal survey\cite{decadal_survey_2010}, its launch is currently expected to occur by the end of 2026.

\textit{Roman} features a 2.4~m primary mirror, similar in size to \textit{Hubble} Space Telescope's, it will observe in the optical and in the NIR. It carries two instruments: the Coronagraph and the Wide Field Instrument (WFI). WFI will have a field of view 100 times larger than \textit{Hubble}'s, making it an outstanding instrument for large surveys, while providing excellent wavelength coverage with 7 bands broad bands from 0.62~$\mu$m to 2.13~$\mu$m, ensuring a a continuous coverage of the rest-frame near ultraviolet (NUV) and the optical at redshift 2 and beyond. The majority of the observation time of its initial five-year mission will be dedicated to Core Community Surveys (CCS). Of particular interest is the High Latitude Wide Area survey, which will be dedicated to large extragalactic fields, aiming to detect around a billion galaxies. In addition to the CCS program, \textit{Roman} will also accommodate competitively selected ``General Astrophysics Surveys'' with the WFI, further expanding the mission's scientific reach, and giving opportunities for combined follow-up observations with PRIMA.

\section{Methodology\label{sec:methodology}}

Our approach revolves around Spectral Energy Distribution (SED) modeling, which we present in Sec.~\ref{sec:SED} and the CIGALE modeling code in particular in Sec.~\ref{sec:cigale}. This allows us to build a simulated synthetic dataset of dusty star-forming galaxies with \textit{Roman} and PRIMA at different redshifts (Sec.~\ref{sec:build-dataset}), and subsequently estimate their properties.

\subsection{Spectral Energy Distribution modeling\label{sec:SED}}
Since the seminal works of Tinsley\cite{tinsley1972a,tinsley1976a}, SED modeling has played an increasingly important role in extragalactic astronomy. The first models were mostly limited to passive stellar populations. However, even such early models proved remarkably powerful when fitted to observations in order to measure physical properties such as the stellar mass of elliptical galaxies, for instance. Over the years, these models have seen great developments, progressively including all the main baryonic physical components of galaxies (stars, ionized gas, dust in absorption and emission, and even active nuclei), expanding their wavelength coverage with some models now extending across the electromagnetic range from the X-rays to the radio, and overall improving their quality and reliability. Combining these models with statistical techniques, such as $\chi^2$ minimization or a Bayesian approach, SED fitting is now the \textit{de facto} method to estimate the main physical properties of galaxies across the universe.

Early popular SED-fitting codes, such as Hyperz\cite{hyperz} or Le Phare\cite{le_phare}, were able to derive redshifts and other physical parameters using only visible and NIR images. New-generation codes like MAGPHYS\cite{magphys}, BAGPIPES\cite{carnall2018a}, or Prospector\cite{leja2017a} are now routinely used for inferring the physical properties of galaxies from the FUV to the FIR. In this work we adopt the Code Investigating Galaxy Emission\cite{cigale} (CIGALE) spectro-photometric modeling code.

Although different codes could have been used for this work, they generally lead to similar results\cite{pacifici2023a} and CIGALE offers the necessary flexibility we need to both generate the dataset of synthetic \textit{Roman} and PRIMA observations and estimate the physical properties from such observations.

\subsection{CIGALE\label{sec:cigale}}
CIGALE is based on energy balance. This means that the energy of the photons absorbed by the dust at UV to NIR wavelengths corresponds exactly to the total energy emitted by the dust. Even though it is in principle necessary to solve the energy transfer equation to compute the exact spectrum of a galaxy, such an approach is resource-intensive. In practice, the energy balance approximation provides excellent results for a fraction of the computing cost. For instance, CIGALE can compute a full multi-wavelengths model in less than a millisecond on a single computing core, while being able to compute many models in parallel on different cores. This allows CIGALE to produce large grids of models in a reasonable amount of time and fit them to observations. The physical properties and their uncertainties are then inferred from the likelihood-weighted means and standard deviations over the entire grid.

As we want to build a realistic synthetic dataset of galaxies observed with PRIMA and \textit{Roman} and measure their properties, it is important that these models account for the main physical components of galaxies. CIGALE adopts a modular approach, with each physical component corresponding to a specific module. The modules are then combined to build the multi-wavelengths model. The most recent version handles, among other things, complex SFH with a variety of stellar population models, nebular emission (recombination lines and continuum emission), the absorption by dust using flexible attenuation curves, the re-emission of the absorbed energy in the MIR and FIR, and the effect of the absorption by the intergalactic medium as photons propagate through the universe.

\subsection{Generation of PRIMA and \textit{Roman} synthetic observations\label{sec:build-dataset}}
\subsubsection{General principle}
As PRIMA and \textit{Roman} are still in preparation, we rely on a synthetic dataset of simulated observations that reflects the depth of the planned extragalactic surveys. Generating such a dataset in great detail can become excruciatingly complex, in particular to reproduce finely the emission spectrum of distant galaxies, whose variations, in particular in the rest-frame MIR where Polycyclic Aromatic Hydrocarbon (PAH) features dominate, remain poorly constrained.

An approach is to rely on numerical simulations and semi-analytical models\cite{bisigello2021a}. This technique can give excellent results for building, for instance, realistic luminosity functions, evaluate the impact of confusion noise at longer wavelengths, etc. However, for the present study there are two main drawbacks. First, at this time we are not aware of any catalog of simulated observations that would consistently model both PRIMA and \textit{Roman} simultaneously. Generating such a dataset from simulations would represent an amount of work that is beyond our current resources. The second limitation is that fitting existing catalogs with CIGALE, we would run the risk of injecting biases in our results that would stem from differences in the basic bricks used to model the galaxies (e.g., different stellar populations or dust models).

To avoid this pitfall and assess the science case considered, we have selected a simple yet potent approach. We modeled with CIGALE a synthetic dataset of dust star-forming galaxies with a stellar mass of $10^{10.0}$, $10^{10.5}$, $10^{11.0}$, and $10^{11.5}$~M$_\odot$ observed with both PRIMA and \textit{Roman} at $z=1.5$, $z=2.0$, and $z=2.5$ (see section \ref{sec:primager_simulations} for a description of how the characteristics of each synthetic galaxy is set). We have selected such a discretization as our goal is not to explore and establish, for instance, constraints on luminosity function, but to assess the quality of the characterization of individual galaxies of different masses at different redshifts. We then fit these simulated observations with CIGALE to estimate the physical properties of each of these galaxies in the same way we would for actual observations.

Our method has the advantage of being fully consistent, eliminating poorly-controlled biases that would be due to differences between the catalog-generation and fitting models, while still taking into account the effect of survey depth and photometric uncertainties. In practice, if we were to use different codes (or even changing the details of the modeling while keeping the same code, for instance using different dust templates) for different steps, we would not be in able to interpret the results with confidence. That is, we would not be able to determine if any difference between the ``true'' and ``estimated'' properties we find would stem from fundamental differences between the models themselves or if they are the consequence of the quality of the observations (wavelength coverage, signal-to-noise ratio) and the characteristics of the objects observed (redshift, stellar mass, SFR, PAH mass fraction qPAH, dust luminosity), which is what we are investigating in this work. However, it is important to keep in mind that exploring synthetic catalogs does not inform us whether we would be able to retrieve the intrinsic parameters from actual observations as the models may prove imperfect to reproduce such observations with sufficient precision and accuracy. In such case, comparing the results obtained using different codes would provide useful information on the uncertainty range stemming from the limits of the models themselves\cite{hunt2019a,pacifici2023a}.

\subsubsection{Building the synthetic PRIMA and \textit{Roman} datasets}\label{sec:primager_simulations}
To build the synthetic dataset, the first step is to determine how the SFH, stellar populations, ionized gas, dust attenuation, and dust emission are modeled. We provide here a brief description of each of these components.

\paragraph{Star Formation History.} Because they are short-lived, the massive stars that dominate the blue end of the electromagnetic spectrum are very sensitive to the SFH\cite{boquien2014a}. To account for the variation of the SFH over different timescales, we model the SFH using the \texttt{sfhdelayedbq} module\cite{ciesla2016a}, which combines a long-term delayed SFH of timescale $\tau_\textrm{main}$ and age $\textrm{Age}_\textrm{main}$ with a shorter-term constant star formation burst or quench that started $\textrm{Age}_\textrm{bq}$ before present and with a relative amplitude $r_\textrm{SFR}$.

\paragraph{Stellar populations.} Having defined the SFH, to generate the stellar spectrum, we use the \texttt{bc03} module, which is based on the Bruzual \& Charlot 2003 single stellar populations\cite{bruzual} with a Chabrier IMF\cite{chabrier2003a}. The combination of the SFH with these models allows to compute the intrinsic stellar spectrum.

\paragraph{Ionized gas.} The \texttt{nebular} module models the emission of gas ionized by Lyman continuum photons emitted by the complex stellar populations created at the previous step. These models include recombination lines, including fine-structure FIR lines such as [OIII] at 88~$\mu$m and [CII] at 158~$\mu$m, as well as continuum emission from free-free, free-bound, and 2-photon processes. The strength of this nebular emission is directly linked to the rate at which high-energy Lyman continuum photons are produced, depending on the ionization parameter $\textrm{log U}$, the gas metallicity and the electron density $\textrm{n}_\textrm{e}$. We assume that the escape fraction of the these photons is zero.

\paragraph{Dust attenuation.} Dust plays a major role in how galaxies appear to us. The physical characteristics and the way dust is distributed great influences how much light is absorbed at different wavelengths\cite{charlot2000a} and how the different components are affected. This leads to variations in the attenuation curves. To account for these variations, we use the flexible \texttt{dustatt\_modified\_CF00} module, which implements the Charlot \& Fall\cite{charlot2000a} model that is implemented as the combination of two power laws: the first one only attenuates radiation from birth clouds (index $\delta_\textrm{BC}$, for stars younger than 10 Myr), and the second one, with a V-band attenuation $\textrm{Av}_\textrm{ISM}$, models the interstellar dust attenuating stellar populations of all ages (index $\delta_\textrm{ISM}$), including the radiation that has already undergone some level of attenuation in the birth clouds. The combination is set through the ratio $\mu$ of the attenuation of the interstellar dust to the total attenuation generated by both the interstellar dust and that in birth clouds.

\paragraph{Dust emission.} To model dust emission, we use the \texttt{dl2014} module, which is based on the Draine \& Li dust emission models\cite{draine2007a,draine2014a}. For our purposes, obtaining reliable estimates from dust emission is crucial as we are especially interested in the dust luminosity and the PAH abundance $\textrm{q}_\textrm{pah}$. It combines two components, the first one modeling the interstellar dust illuminated by a radiation field $U_{min}$ and the second one with a mass fraction $\gamma$ that models dust associated to star-forming regions and that is illuminated with a range of radiation field intensities following a power law : $dM/dU\propto U^{-\alpha}$.

\paragraph{Drawing the physical properties of each galaxy.}
Each of these modules has a number of parameters. To build our dataset, we generate a unique set of parameters to each galaxy. For simplicity, we set some parameters to fixed values for the whole sample while others are free. To ensure we have a reasonably diverse set of galaxies, the free parameters are randomly sampled from a uniform distribution that can be discrete or continuous, depending on their nature. In Table~\ref{tab:cigale_param}, we indicate the main parameters and the values or sampling associated with each of the CIGALE modules. We refer to the CIGALE documentation for the detailed description of each parameter\cite{ciesla2016a,cigale}.

\begin{table*}[!htbp]
    \centering
    \begin{tabular}{lll}
        \hline
         Module & Parameters & Values  \\
         \hline
         \hline
         \texttt{sfhdelayedbq} & $\tau_\textrm{main}$ (Myr) & $\left[4000-13000\right]$\\
                               & $\textrm{Age}_\textrm{main}$ (Myr) &  Age of the universe - 300\\
                               & $\textrm{Age}_\textrm{bq}$ (Myr) &  $\left[60-160\right]$\\
                               & $r_\textrm{SFR}$ & $\left[0.6-12\right]$ \\
         \hline
         \texttt{bc03} & Metallicity & $0.02$\\
                       & IMF & Chabrier 2003\\
          \hline
          \texttt{nebular} & $\textrm{log U}$ & $-3$ \\
                           & $\textrm{Z}_\textrm{gas}$ & $0.02$ \\
                           & $\textrm{n}_\textrm{e}$ (cm$^\textrm{-3}$) & $100$ \\
          \hline
          \texttt{dustatt\_modified\_CF00} & $\textrm{Av}_\textrm{ISM}$ (mag) & $\left[0.5-2.0\right]$ \\
                                           & $\mu$ & $\left[0.25-0.50\right]$ \\
                                           & $\delta_\textrm{BC}$ & $-0.7$ \\
                                           & $\delta_\textrm{ISM}$ & $-1.3$ \\
          \hline
          \texttt{dl2014} & $\textrm{q}_\textrm{pah}$ & $\left[0.47-7.32\right]$ \\
                          & $U_{min}$ & $30, 35, 40, 45$ \\
                          & $\alpha$ & $\left[1.6-2.4\right]$ \\
                          & $\log \gamma$ & $\left[-2.0-0.5\right]$ \\
          \hline
    \end{tabular}
    \caption{CIGALE parameters used to model generate the synthetic catalog. The values given between brackets represent an uniform distribution between the two bounds.}
    \label{tab:cigale_param}
\end{table*}

\paragraph{Generating synthetic observations.}
With this configuration, we draw 1000 galaxies in each redshift bin ($z=1.5$, $2.0$, and $2.5$). In the following step, we use this catalog of input parameters to build the corresponding models with CIGALE, determining in particular their fluxes in the PRIMA and \textit{Roman} bands.

The PRIMAger Hyperspectral Imager (PHI) utilizes linear variable filters (LVF) for hyperspectral imaging across two bands\cite{Ciesla25}, PHI1 and PHI2 that covers a total range from 24 to 84~$\mu$m. In our experiment, the PHI observations are simulated with 48 filters built with a Lorentzian profile assuming $R=10$, which originates from the team developing the linear variable filter at SRON, as manufacturing constraints and results. PRIMAger will also be able to image with polarimetry via four broad band channels, from PPI1 to PPI4. The filter transmission from each band of \textit{Roman}’s WFI has been retrieved from the Spanish Virtual Observatory filter service website (\url{http://svo2.cab.inta-csic.es/svo/theory/fps3/index.php}).

At this stage, CIGALE provides us with the physical properties and the corresponding fluxes. However, these properties and fluxes are all determined in an exact sense, to the extent that the model faithfully reproduces such galaxies. For a realistic simulation of the observations of distant galaxies with PRIMA and \textit{Roman}, we need to take into account observational noise stemming from the depth of the surveys and photometric measurements. This is necessary for the computation of the photometric uncertainties and to simulate the measured fluxes.

To compute the uncertainties we need three ingredients: the absolute scale of the fluxes, the characteristics of the survey, and a general photometric error term. The first term allows to examine galaxies of different masses. As our objective is not to assess galaxies at the population level but rather to characterize individual galaxies, we scale the galaxies at each redshift to four mass bins: $10^{10.0}$, $10^{10.5}$, $10^{11.0}$, and $10^{11.5}$~M$_\odot$, quadrupling the sample size from 1000 to 4000 galaxies per redshift. For the second ingredient, we need to considerate the different surveys for these two instruments. While there is no guarantee that surveys will indeed overlap, we aim to examine here the advantages of such an overlap. We therefore adopt PRIMA's Deep Survey\cite{bisigello2024a} and \textit{Roman}'s High Latitude Survey (HLS)\cite{akeson2019a}. Due to the observations at fairly long wavelengths, PRIMA's uncertainties have a strong contribution originating from the confusion noise rather than the depth of the survey itself. For both the survey sensitivity and the confusion noise, we adopt the ``conservative'' values listed in Table~1 of Bisigello et al. (2024)\cite{bisigello2024a}. Given that we consider 24 filters in each of the PHI1 and PHI2 bands rather than 6 as was assumed in the Deep Survey paper, we degrade the sensitivity of the survey by a factor 2 for computing the photometric noise in individual filters. Summing the sensitivity noise and the confusion noise in quadrature, this leads to 1-$\sigma$ uncertainties ranging from 35~$\mu$Jy for the PHI1 band at 24~$\mu$m to 222~$\mu$Jy for the PHI2 band at 84~$\mu$m. The confusion noise increases rapidly in the PPI bands and we consider only the PPI1 and PPI2 bands, with a combined uncertainty of 340~$\mu$Jy and 821~$\mu$Jy respectively. These uncertainties are taken as uncorrelated. However, an important aspect with actual PRIMAger observations is that uncertainties between different filters will probably show a certain degree of correlation due to the nature of the filters. To our knowledge no detailed noise model is currently available and their inclusion in the fitting is left to future work. As for \textit{Roman}'s HLS, it is a four-band survey with a 1-$\sigma$ sensitivity of approximately 13~$\mu$Jy (F106, F129, and F158 bands) to  24~$\mu$Jy (F184 band). Finally, we consider a general photometric error source corresponding to 5\% of the intrinsic flux which encompasses errors in the photometric measurements themselves (e.g. due to imperfect aperture corrections, etc.) and uncertainties in the models. This component, which scales with the galaxy's mass, is also summed in quadrature. A summary of the selected PRIMA and \textit{Roman} filters is presented in Table~\ref{tab:prima_roman_filter_selection}.

\begin{table}[!htbp]
\centering 
    \begin{tabular}{lcrr}
    \hline
     Instrument & Bands & Wavelength & 1-$\sigma$ sensitivity \\
     \hline
     \hline
     WFI      & F106 & 1.06~$\mu$m & 0.013 $\mu$Jy\\
              & F129 & 1.29~$\mu$m & 0.013 $\mu$Jy\\
              & F158 & 1.58~$\mu$m & 0.013 $\mu$Jy\\
              & F184 & 1.84~$\mu$m & 0.024 $\mu$Jy\\\hline
     PRIMAger & PHI1     & 24--45~$\mu$m & 35--90 $\mu$Jy\\
              & PHI2     & 45--84~$\mu$m & 90--222 $\mu$Jy\\
              & PPI1     & 92~$\mu$m & 340 $\mu$Jy\\
              & PPI2     & 126~$\mu$m & 812 $\mu$Jy\\\hline
    \end{tabular}
    \caption{Selected \textit{Roman} and PRIMA filters for this study.}
    \label{tab:prima_roman_filter_selection}
\end{table}

To build the final catalog of synthetic galaxies, we assign to each galaxy the fluxes and uncertainties computed above. Then, for each band we disturb the flux. For this, we draw a new flux from a Gaussian distribution that is centered on the intrinsic flux and with a 1-$\sigma$ standard deviation corresponding to the aforementioned uncertainty. With this processing carried out, we have at our disposal a synthetic catalog of simulated observations of galaxies for which we know the ground truth and which form the basis of our analysis. For some galaxies, the randomly-drawn noise can lead to very low or even negative fluxes, which is statistically expected. Even though CIGALE handles upper limits natively, in this case we have chosen to leave these low fluxes as-is as they naturally have a low relative weight in the computation of the likelihood.

A last important point in the constitution of this catalog, is that we also have largely neglected some aspects of the data processing that will be important with actual observations. A major step is that of deblending between close sources. For this work, we have assumed that deblending is carried-out beforehand. Important efforts are underway by the PRIMA team to develop deblending algorithms and a first analysis is available\cite{donnellan,bethermin2024a}.

\section{Results and discussion}\label{sec:results}

\begin{figure}[!htb]
    \centering
    \includegraphics[width=\textwidth]{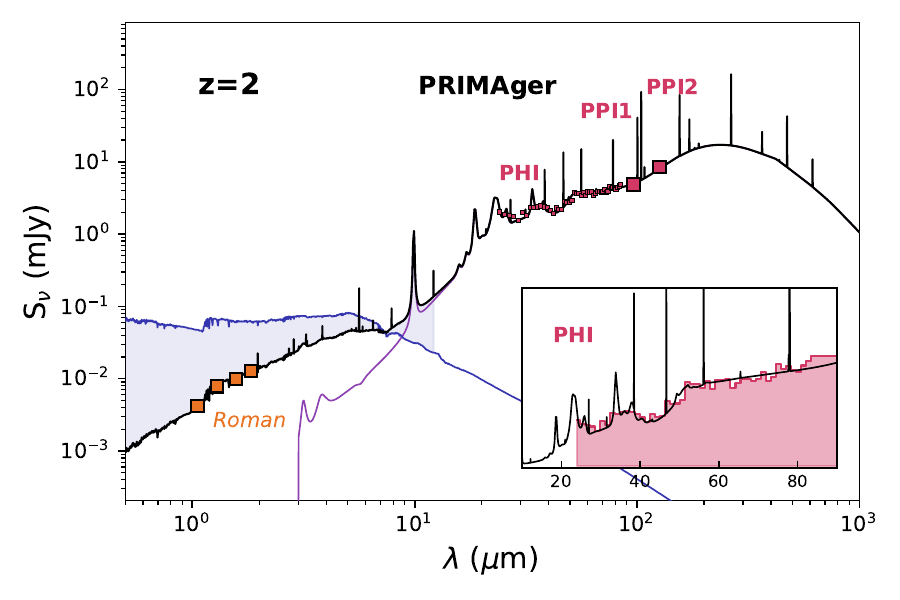}
    \caption{Typical best-fit of synthetic galaxy with the PRIMA and \textit{Roman} bands (black). The model includes stellar populations (unattenuated in blue) and dust in emission (purple). The synthetic observations are given as orange squares for \textit{Roman} observations and red squares for PRIMA, with the error bars showing the 3-$\sigma$ uncertainties. The inset panel zooms-in the PHI observations between 24 and 84\,$\mu$m.}
    \label{fig:fit_prima_roman}
\end{figure}

To model the synthetic catalog built in the previous section, we use CIGALE once more, fitting these observations with a reasonably extensive grid of 42768000 models that is given in Table~\ref{tab:cigale_model}.
\begin{table*}[!ht]
    \centering
    \begin{tabular}{lll}
        \hline
         Module & Parameters & Values  \\
         \hline
         \hline
         \texttt{sfhdelayedbq} & $\tau_\textrm{main}$ (Myr) & 3000, 6000, 9000, 15000\\
                               & $\textrm{Age}_\textrm{main}$ (Myr) &  Age of the universe - 300\\
                               & $\textrm{Age}_\textrm{bq}$ (Myr) &  50, 100, 200\\
                               & $r_\textrm{SFR}$ & 0.5, 1.0, 1.5, 2.0, 3.0, 4.0, 5.0, 10.0, 15.0 \\
         \hline
         \texttt{bc03} & Metallicity & $0.02$\\
                       & IMF & Chabrier 2003\\
          \hline
          \texttt{nebular} & $\textrm{log U}$ & -3 \\
                           & $\textrm{Z}_\textrm{gas}$ & 0.02 \\
                           & $\textrm{n}_\textrm{e}$ (cm$^\textrm{-3}$) & 100 \\
          \hline
          \texttt{dustatt\_modified\_CF00} & $\textrm{Av}_\textrm{ISM}$ (mag) & 12 values from 0.25 to 2.25 \\
                                           & $\mu$ & 0.2, 0.3, 0.4, 0.5, 0.6 \\
                                           & $\delta_\textrm{BC}$ & $-0.7$ \\
                                           & $\delta_\textrm{ISM}$ & $-1.3$ \\
          \hline
          \texttt{dl2014} & $\textrm{q}_\textrm{pah}$ & 15 values from 0.47 to 7.32 \\
                          & $U_{min}$ & 30, 35, 40, 45 \\
                          & $\alpha$ & 1.5, 1.6, 1.7, 1.8, 1.9, 2.0, 2.1, 2.2, 2.3, 2.4, 2.5 \\
                          & $\log \gamma$ & 10 values from -2.0 to -0.5 \\
          \hline
    \end{tabular}
    \caption{CIGALE parameters used to model generate the synthetic catalog.}
    \label{tab:cigale_model}
\end{table*}
This grid is recomputed separately for each redshift bin. We note that even though CIGALE has the capability of fitting objects of unknown redshift, we assume that the galaxies that will be modeled for such a survey will already have a photometric or a spectroscopic redshift readily available. We present an example of a typical best-fit using the PRIMA and \textit{Roman} bands in Fig.~\ref{fig:fit_prima_roman} for a $10^{11.0}$~M$_\odot$ galaxy at $z=2$.

Even though the noise is fairly high in individual PHI filters, the large number of bands offered by PRIMAger still allows to trace the dust emission in detail. Compared to actual observations, another factor leading to lower $\chi^2$ is that the models we employ are consistent for both the creation of the synthetic catalog and for the estimations of the physical properties from the fits. As mentioned earlier, our technique naturally eliminates the poorly-controlled biases that originate from imperfections in the various model components (stars, dust, and gas). This means our results are mostly sensitive to the set of bands and the observational noise characteristics set by the observatory and survey parameters.

\begin{figure}[!htb]
    \centering
    \includegraphics[width=\textwidth]{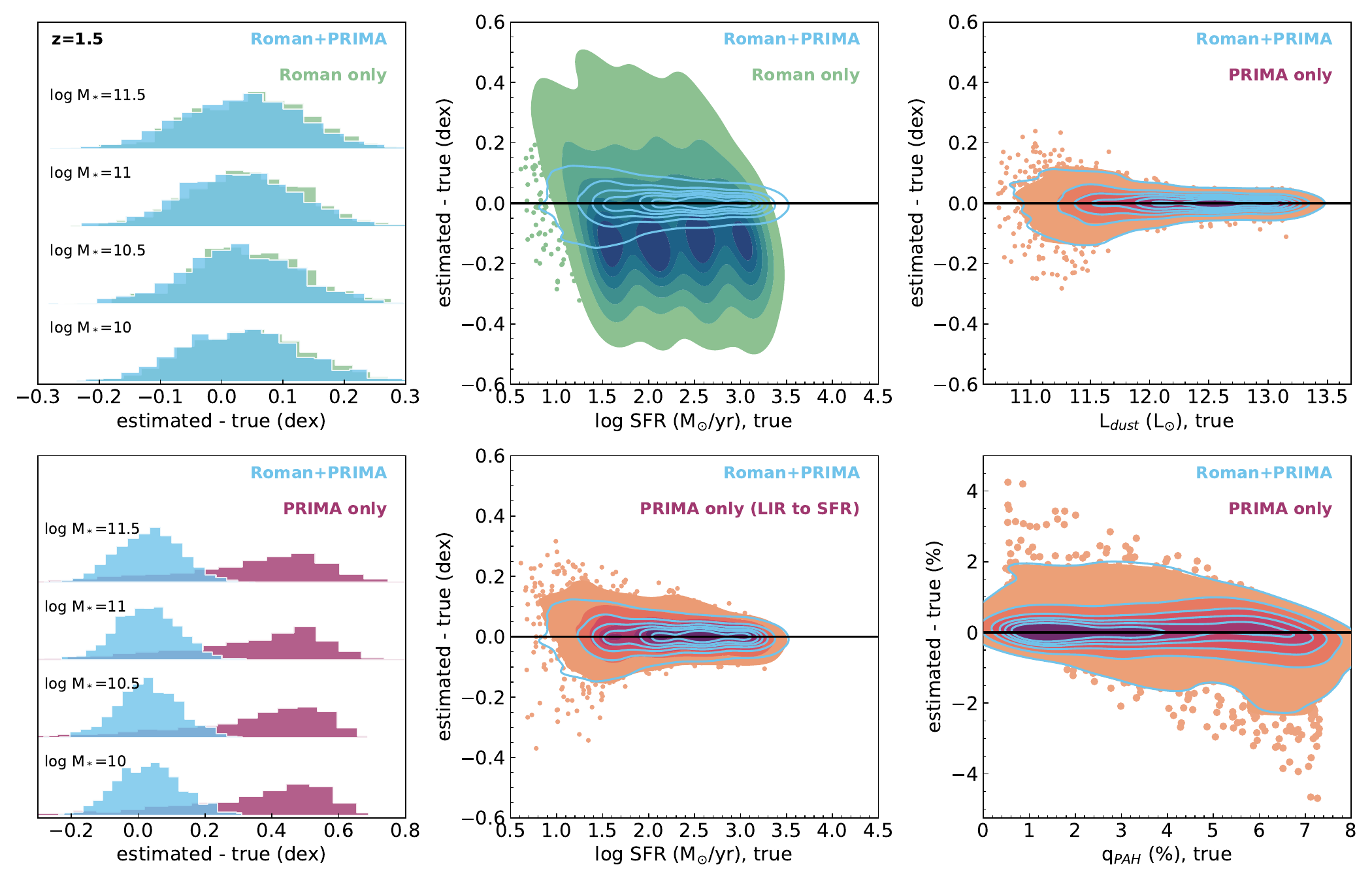}
    \caption{Comparison between the true and estimated valued at $z=1.5$ for the following physical properties (from left to right and top to bottom): stellar masses, star formation rate, dust luminosity, and PAH fraction. In these panels, the results obtained from the PRIMA+\textit{Roman} combination of filters are shown in light blue while results obtained using only \textit{Roman} or PRIMA are indicated in green and orange, respectively. }
    \label{fig:mock1p5}
\end{figure}

\begin{figure}[!htb]
    \centering
    \includegraphics[width=\textwidth]{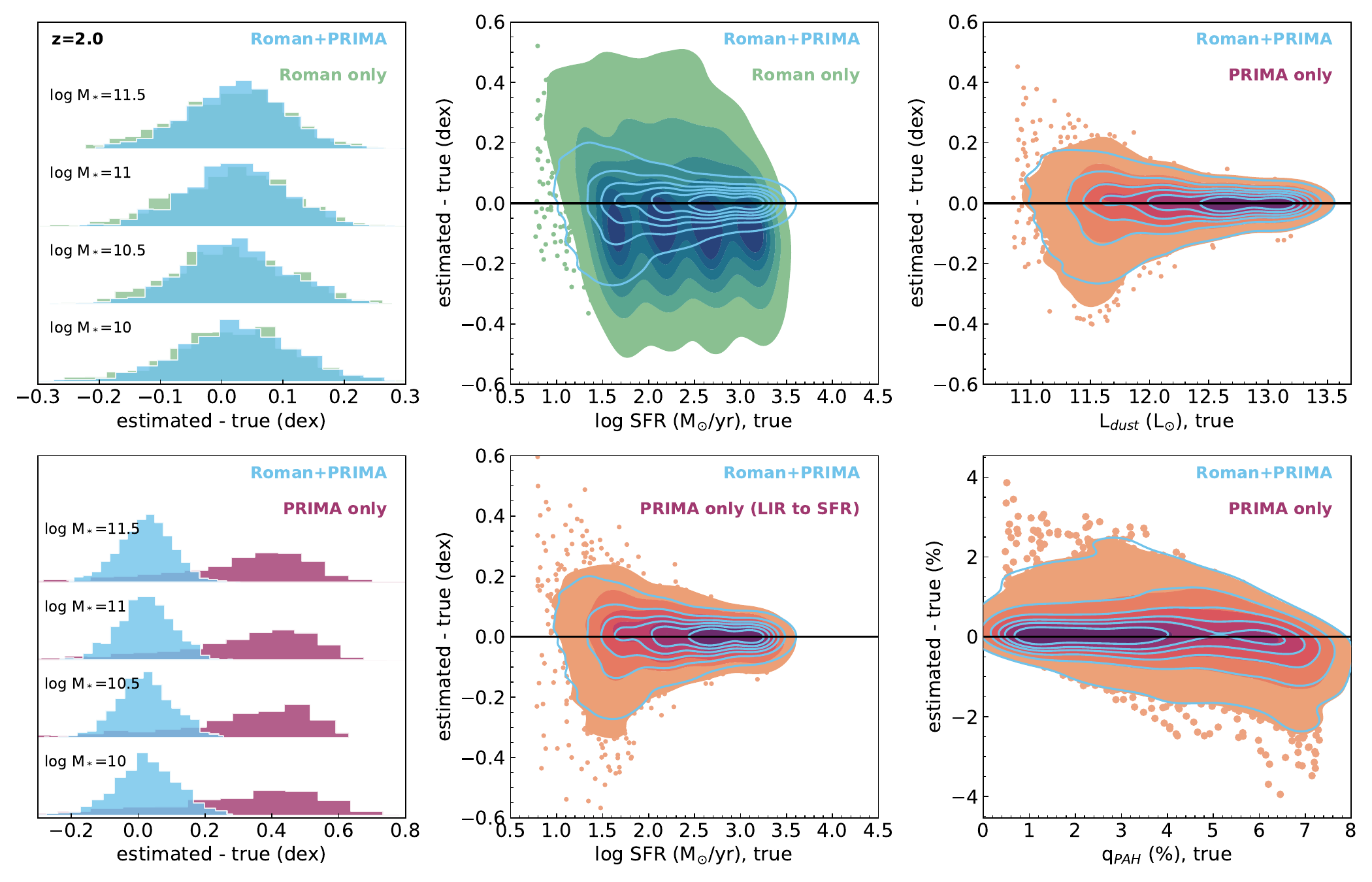}
    \caption{Same as Fig.~\ref{fig:mock1p5} but at $z=2.0$.}
    \label{fig:mock2p0}
\end{figure}

\begin{figure}[!htb]
    \centering
    \includegraphics[width=\textwidth]{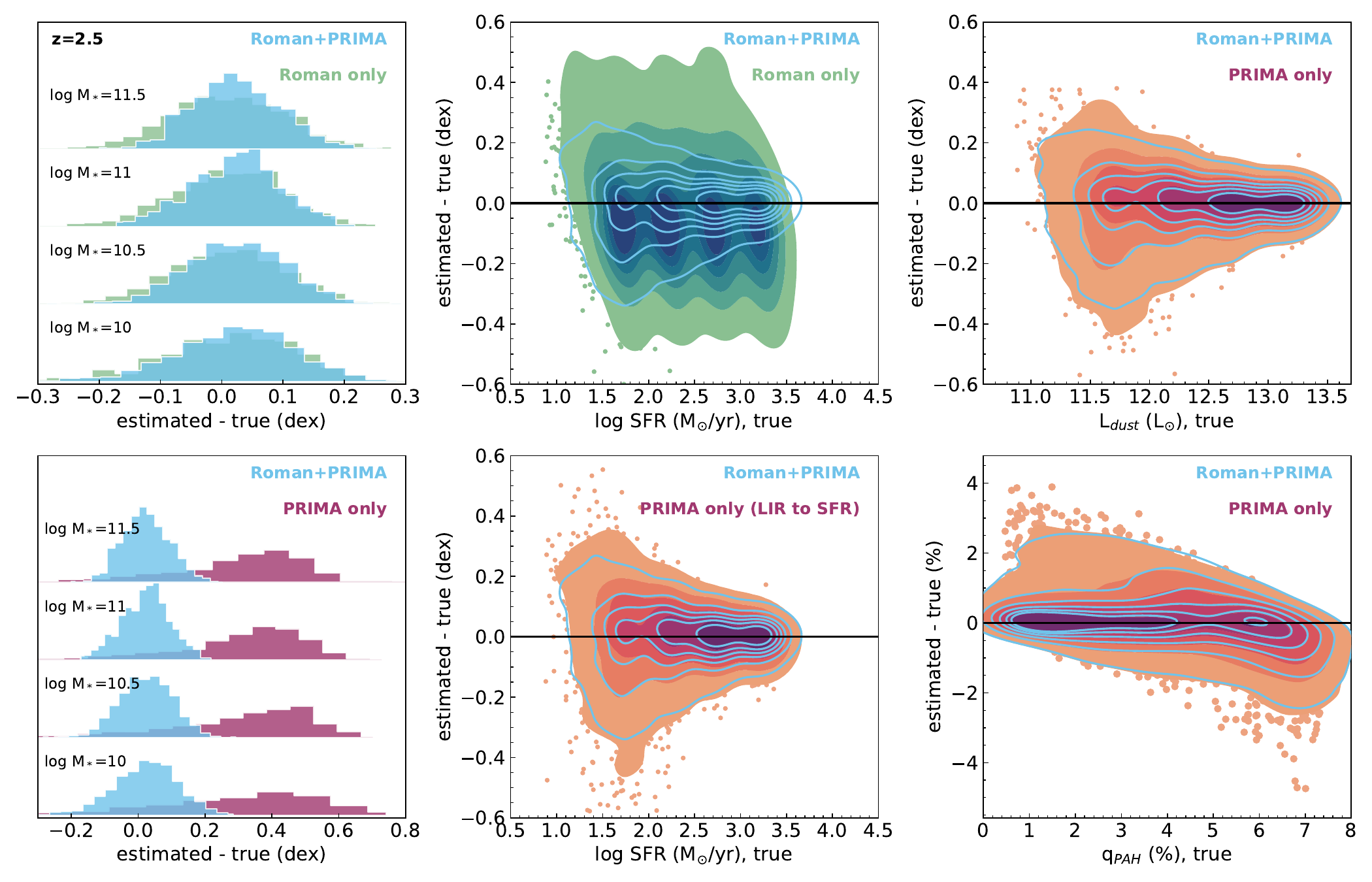}
    \caption{Same as Fig.~\ref{fig:mock1p5} but at $z=2.5$.}
    \label{fig:mock2p5}
\end{figure}

However, having a good fit does not necessarily mean that the physical properties are well estimated. The best-fit only provides us with a very limited view on possible solutions. There could be solutions with marginally higher $\chi^2$, that would have been selected with another realization of the observations of the same galaxy. An approach would be to generate multiple realizations of the observations of each galaxy and establish the mean and the standard-deviation of the best-fit estimations. However, this would be highly resource-intensive. Rather, to account for this, we prefer to rely on a Bayesian approach for parameter estimation. CIGALE provides such an estimate through the likelihood-weighted mean over all the models. At the same time, this approach also provides us with a natural estimate of the corresponding uncertainties through the likelihood-weighted standard deviation. We present in Fig.~\ref{fig:mock1p5}, Fig.~\ref{fig:mock2p0}, and Fig.~\ref{fig:mock2p5}, the comparison between the true values and the estimates of 1) the stellar mass, 2) the star formation rate, 3) the total dust luminosity, and 4) $q_{PAH}$ at $z=1.5$, $z=2.0$, and $z=2.5$, respectively. Furthermore, we test these estimates using either PRIMA only \textit{Roman} only, PRIMA+\textit{Roman} set of filters.

We compare the physical parameters extracted from the PRIMA and \textit{Roman} synthetic observations to the ground-truth values used to generate the models. The medians and the standard deviations of the differences between the estimated and exact values obtained with PRIMA only, \textit{Roman} only, and when combining \textit{Roman} and PRIMA are presented in Table~\ref{tab:rel_err_parameters_redshift}.

\begin{table}[!htbp]
    \centering
    \begin{tabular}{lc|ccc}
         \hline
         Parameter & z & PRIMA & \textit{Roman} & PRIMA+\textit{Roman} \\
         \hline
         \hline
         Stellar Mass        & 1.5 & $0.39\pm0.23$ & $0.04\pm0.09$ & $0.03\pm0.10$ \\
                             & 2.0 & $0.35\pm0.23$ & $0.02\pm0.10$ & $0.02\pm0.09$ \\
                             & 2.5 & $0.35\pm0.23$ & $0.01\pm0.10$ & $0.03\pm0.08$ \\
         \hline
         Star Formation Rate & 1.5 & $-0.01\pm0.06$ & $-0.09\pm0.20$ & $0.00\pm0.05$ \\
                             & 2.0 & $-0.01\pm0.10$ & $-0.05\pm0.21$ & $0.00\pm0.08$ \\
                             & 2.5 & $0.00\pm0.13$ & $-0.05\pm0.21$ & $0.00\pm0.10$  \\
         \hline
         Dust Luminosity     & 1.5 & $0.00\pm0.04$ & $-0.10\pm0.20$ & $0.00\pm0.04$ \\
                             & 2.0 & $0.00\pm0.09$ & $-0.06\pm0.21$ & $0.00\pm0.21$ \\
                             & 2.5 & $0.00\pm0.13$ & $-0.06\pm0.21$ & $0.00\pm0.10$ \\
         \hline
         PAH fraction        & 1.5 & $0.00\pm0.74$ & $0.02\pm2.03$ & $0.01\pm0.73$  \\
                             & 2.0 & $0.02\pm0.78$ & $0.16\pm1.95$ & $0.04\pm0.77$  \\
                             & 2.5 & $0.00\pm0.91$ & $-0.04\pm1.98$ & $0.03\pm0.87$ \\\hline
    \end{tabular}
    \caption{Medians and standard deviations of the difference between the estimated and true values for PRIMA only, \textit{Roman} only, and PRIMA+\textit{Roman}. The stellar mass, star formation rate, and dust luminosity are given in dex and the PAH fraction is the difference in percent.}
    \label{tab:rel_err_parameters_redshift}
\end{table}

In general, at the three redshifts we investigate in this work, the stellar mass is well recovered thanks to the \textit{Roman} wavelength coverage with a median difference between the exact and estimated values lower than 0.03\,dex and a dispersion lower than 0.10~dex at all redshifts. With its wavelength coverage, PRIMA cannot recover the stellar mass as the rest-frame near-IR is not probed at all. The median difference of $\sim$0.4~dex on the stellar mass, which we can also observe in the bottom-left panels of Figs.~\ref{fig:mock1p5} to \ref{fig:mock2p5} when using only PRIMA data only shows this parameter is mostly unconstrained. The amplitude of the offset is dependent on the details of the modeling and should not be relied upon. In effect, when relying exclusively on PRIMA observations, the stellar mass is estimated indirectly from the dust mass and it thus unreliable and possibly systematically biased.

As redshift increases, \textit{Roman} bands progressively become sensitive to rest-frame blue and even NUV wavelengths, therefore, regarding the SFR, \textit{Roman} alone recovers the properties with a median value decreasing from $-0.09\pm0.20$\,dex to $-0.05\pm0.21$\,dex from $z=1.5$ to $z=2.5$. Adding PRIMA observations allows to improve the accuracy on the SFR estimates to $0.00\pm0.10$\,dex at $z=2.5$. This result highlights the strong complementarity of \textit{Roman} and PRIMA regarding one of the key physical properties of galaxies. Based on the estimates of the dust luminosity, PRIMA only can be used to estimate the SFR from a direct conversion of the L$_{\rm dust}$. This dust-based SFR results in estimates consistent with the combination of PRIMA+\textit{Roman} indicating than PRIMA alone will be able to provide good SFR measurements. This is the case in particular as the galaxies in our sample tend to be quite dust-rich (A$_{\rm V}$ from 0.5 to 2~mag for the ISM component, see Table~\ref{tab:cigale_param}), which favors dust estimations with PRIMA at the expense of \textit{Roman}. In effect, the difference between the exact and estimated SFR with only PRIMA data shows that only a small fraction of the SFR is not recovered and this only at the lowest attenuation of our sample.

Finally, we also test how well we recover of the PAH fraction using only PRIMA or PRIMA+\textit{Roman} with the goal to understand if a broader set of information on the continuum, in terms of wavelength coverage, would help estimating the contribution of the PAH. There is no improvement on the estimate of q$_{\rm PAH}$ by adding any constraint on the continuum at shorter wavelengths. However, we find that PRIMA alone is sufficient to reliably recover $q_{PAH}$ to a reasonable degree ($<$0.04$\pm$0.87\%) over this wavelength range. This estimate is consistent with the results presented in Bisigello et al. (2024)\cite{bisigello2024a} who found a dispersion of 0.9\% for $q_{PAH}$, with a similar method using CIGALE as well.

\section{Conclusion \label{sec:conclusion}} 
The extragalactic surveys that will be carried out with the PRIMA and \textit{Roman} observatories will provide us with an exquisite view of the stellar populations and dust emission in distant galaxies. In this article we have explored the potential synergies of these surveys to estimate in particular four physical properties: the star formation rate, the stellar mass, the dust luminosity, and the PAH mass fraction. For this, we have generated a catalog of synthetic galaxies at redshifts 1.5, 2.0, and 2.5, simulating the observations of the typical PRIMA and \textit{Roman} extragalactic surveys for galaxies in four mass bins of $10^{10.0}$, $10^{10.5}$, $10^{11.0}$, and $10^{11.5}$~M$_\odot$.

We found that on their own, \textit{Roman} observations allow to estimate the stellar mass and that PRIMA observations allow to estimate well the SFR from a conversion of the dust luminosity and $q_{PAH}$. However, the combination of \textit{Roman} and PRIMA reduces the dispersion on the SFR. In conclusion, the combination of these large surveys will be a boon to measure the fundamental properties of galaxies at intermediate redshift thanks to their sensitivity and extent.

\appendix

\subsection*{Disclosures}
The authors declare that there are no financial interests, commercial affiliations, or other potential conflicts of interest that could have influenced the objectivity of this research or the writing of this paper.

\subsection* {Code, Data, and Materials Availability}
The scripts to generate the synthetic data used in this manuscript and the data themselves are available upon simple request to the corresponding author. The CIGALE code is freely available at \url{https://cigale.lam.fr/}.

\subsection* {Acknowledgments}
MB gratefully acknowledges support from the ANID BASAL project FB210003. This work was supported by the French government through the France 2030 investment plan managed by the National Research Agency (ANR), as part of the Initiative of Excellence of Université Côte d'Azur under reference number ANR-15-IDEX-01.
LC acknowledges support from the french government under the France 2030 investment plan, as part of the Initiative d’Excellence d’Aix-Marseille Université – A*MIDEX AMX-22-RE-AB-101.
This work was supported by CNES, focused on \textit{Roman} and on PRIMA.
This research has made use of the SVO Filter Profile Service "Carlos Rodrigo", funded by MCIN/AEI/10.13039/501100011033/ through grant PID2020-112949GB-I00.

\bibliography{report}
\bibliographystyle{spiejour}

\end{spacing}
\end{document}